\documentclass[aps,pre,preprint]{revtex4-1}
\usepackage{graphics,amsmath,graphicx,color}
\begin{document}

\title{The path to fracture in granular flows: dynamics of contact networks }


\author{M. Herrera}

\author{S. McCarthy}
\author{S. Slotterback}
\author{E. Cephas}
\author{W. Losert}
\author{M. Girvan}
\email[Correspondence to M. Girvan: ]{girvan@umd.edu}
\affiliation{Department of Physics, and IREAP, University of Maryland, College Park, Maryland, USA}

\definecolor{red}{rgb}{1,0,0}

\date{\today}

\begin{abstract}
Capturing the dynamics of granular flows at intermediate length scales can often be difficult.  We propose studying the dynamics of contact networks as a new tool to study fracture at intermediate scales.  Using experimental 3D flow fields with particle scale resolution, we calculate the time evolving broken-links network and find that a giant component of this network is formed as shear is applied to this system.  We implement a model of link breakages where the probability of a link breaking is proportional to the average rate of longitudinal strain (elongation) in the direction of the edge and find that the model demonstrates qualitative agreement with the data when studying the onset of the giant component.  We note, however, that the broken-links network formed in the model is  less clustered than our empirical observations, indicating that the model reflects less localized breakage events and does not fully capture the dynamics of the granular flow.
 
\begin{description}
\item[PACS numbers]
45.70.-n, 47.57.Gc, 64.60.aq, 64.60.ah
\end{description}
\end{abstract}


\maketitle

\section{Introduction}
Forces in dense granular materials are transmitted through an intricate network of particle contacts.  In jammed granular matter, this contact network is able to support stresses and strains up to a yield stress.  However, under large enough stress or strain, the material will yield and deform, as particle contacts fail.  This failure of the jammed state gives way to flow.  The reproducible nature of many flow fields  \cite{Aharonov2002,Mueth2000} of granular matter indicates that the particle contacts fail as a continuum.  However, it is not possible to fully leave the particle scale: shear band widths can be as small as five particle diameters, so unlike in solid mechanics or fluid dynamics, the scale of gradients in velocity is not well separated from the particle scale.  Therefore continuum equations, for example the velocity field, cannot be fully justified.   

The dynamics of dense granular materials are also not amenable to a purely single particle view.  Indeed, the physics of force networks as well as dynamics near jamming highlight the importance of the intermediate scale.  However, actually measuring this intermediate scale is challenging:  under shear strain, force correlations can be found for particles more than 10 particle diameters apart \cite{Majmudar2005}, and cooperative structures such as force loops \cite{Ball2002} have been found.  The force correlations in the direction of the principal stress axis approximately follow a power law.  String-like dynamics in which strings of particles are observed to move cooperatively do not exhibit a preferred length scale but rather a wide, power-law like  distribution of lengths \cite{VollmayerLee2006}.  Elucidating collective dynamics in flows is further complicated since string-like dynamics are defined relative to a random fluctuating background, not relative to an overall flow.  

This intermediate scale is however ideal for observation through the lens of network theory, and we propose to utilize such an approach to study the onset of fracture in 3D structure.  The heavy-tailed characteristics of both forces and string-like dynamics suggests no characteristic length scale exists.  Instead of attempting to discover a characteristic length scale, our analysis produces the characteristic deformation of the whole system over which the contact network is substantially altered.

Surprisingly little is known about the arrangement of grain contacts in a 3D jammed state in experiments, since grains are not transparent to standard imaging methods. Numerous studies have characterized the force network at the boundaries of 3D packings, for example Mueth et. al. used carbon paper to measure forces on the boundary of a cylindrical cell filled with glass beads and subjected to uniaxial compression \cite{Mueth1998}. Techniques for obtaining the interior contact network include freezing the arrangement with glue and cutting slices through the material or layer-by-layer removal of the grains, but these are destructive methods and are not suitable for investigations of system dynamics.

Very recently, complex network methods have been used to analyze the computationally derived force network present in a granular material \cite{Arevalo2010,Tordesillas2010,Walker2010}.  The focus of these studies is on the characterization of the instantaneous contacts and forces in the system. In this paper we introduce the time-evolving of the broken-links network as a new way to characterize particle rearrangement events in our experimental 3D granular flow.  We utilize the language of percolation on networks to describe the dynamics of the network \cite{Newman1999,Callaway2001}.  This approach provides interesting results on the intermediate length and time scales that are difficult to capture with continuum or particle based approaches.  The application of network theoretic concepts to this broken link network provides new insights at this intermediate length (and time) scale.  Additionally, we implement a model of fracture  based on the average rate of longitudinal strain and compare its results to our experimental data.

\section{3D Shear Flows: Imaging and Processing}  
\begin{figure}[!h]
\begin{center}
\includegraphics*[height=.6\textwidth,angle=0,clip]{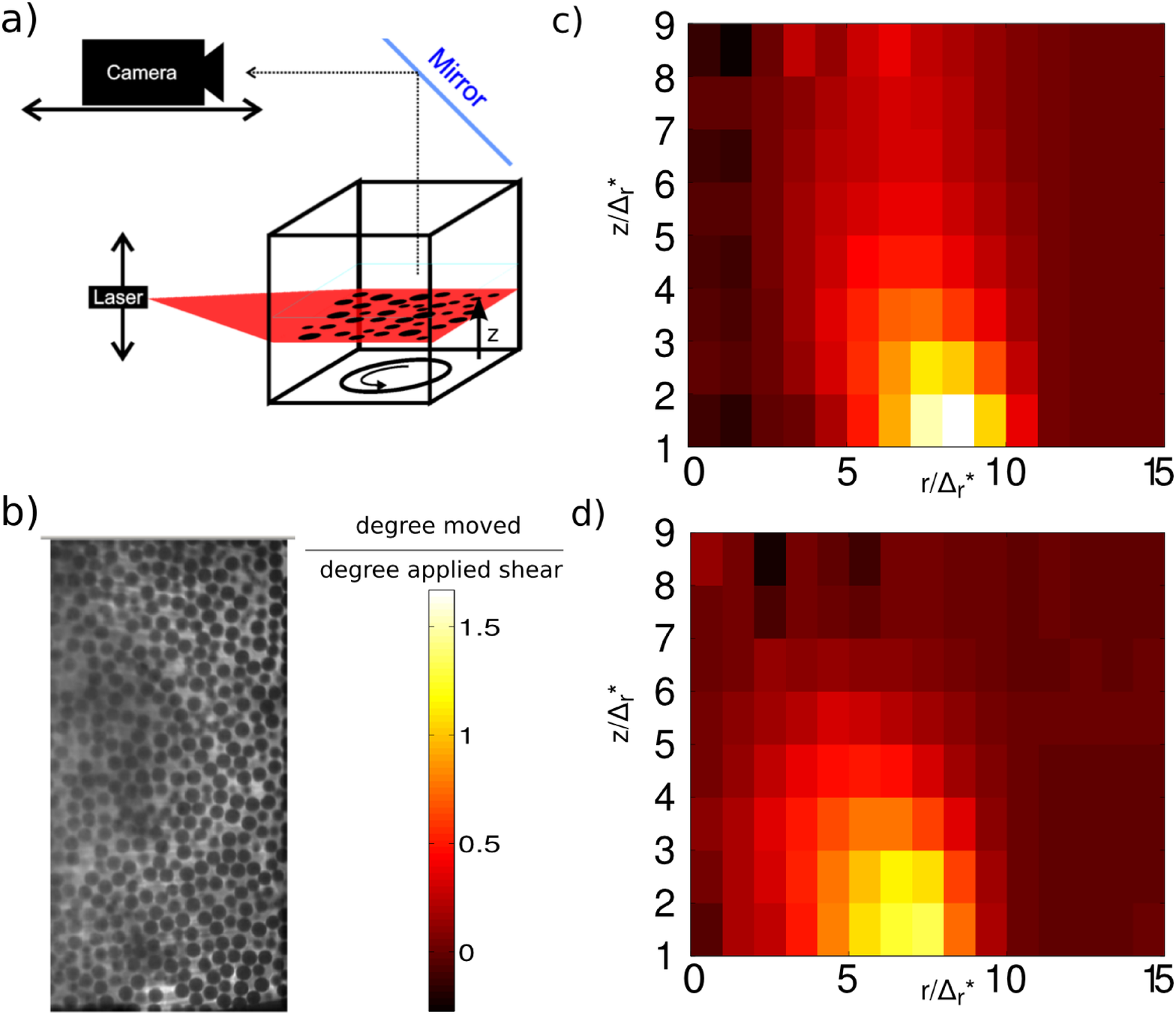}
\caption{(Color Online) (a) Schematic of our experimental setup, a 3D granular imaging method using a laser sheet to illuminate cross sections of the index matched, laser dyed shear cell fluid. (b) An example of an illuminated cross section of the granular flow.  (c) $\frac{1}{\Omega}  r \frac{\partial\omega}{\partial r} $ and (d)  $\frac{1}{\Omega}\ r \frac{\partial\omega}{\partial z} $ the $r$ and $z$ components of the numeric angular velocity gradient.  The region of large gradients corresponds to an area close to the disk edge.  Distances are normalized by the particle contact cutoff $\Delta_r^*=24.5$ (see Figure \ref{palpha}).}
\label{experiment}
\end{center}
\end{figure}

We determine the three dimensional microstructure of a jammed granular material using a laser sheet scanning approach we have described previously \cite{Slotterback2008} which was adapted to study granular shear flows.  The split-cell geometry  as described in \cite{Dijksman2010} consists of a $(15$cm$)^3$ cell with  a $9$cm diameter disk embedded in the bottom.  The cell contains 5mm diameter acrylic beads immersed in an index matched triton and laser dye (Nile Blue 690 Perchlorate) solution, and is filled to a height of 10 particle diameters. The experiment is schematized in Figure \ref{experiment}a.  
Figure \ref{experiment}b displays an example of an illuminated cross section of the granular flow.  We shear the system at a constant rate $\Omega=1.05 \times 10^{-3}$ rad/s, pausing in three degree increments to recover the proximity network as a function of shear (and therefore time).   By scanning a laser, we are able to image slightly more than half the shear cell and reconstruct the individual particle positions in 3D as a function of time.  We bin particle velocities in both $r$ (the radial distance from the center of the disk in cylindrical coordinates) and $z$ (the height above the disk) and compute the average angular velocity $\omega$, and its derivatives with respect to $r$ and $z$ shown in Figures \ref{experiment}c and \ref{experiment}d.  These quantities are used to develop a simple model for link breakage events due to longitudinal strain.  The precision of the particle positions in the shear cell is approximately $2\%$ of the radial cutoff defined below.  

\begin{figure}[!h]
\begin{center}
\includegraphics*[height=.44\textwidth,angle=0,clip]{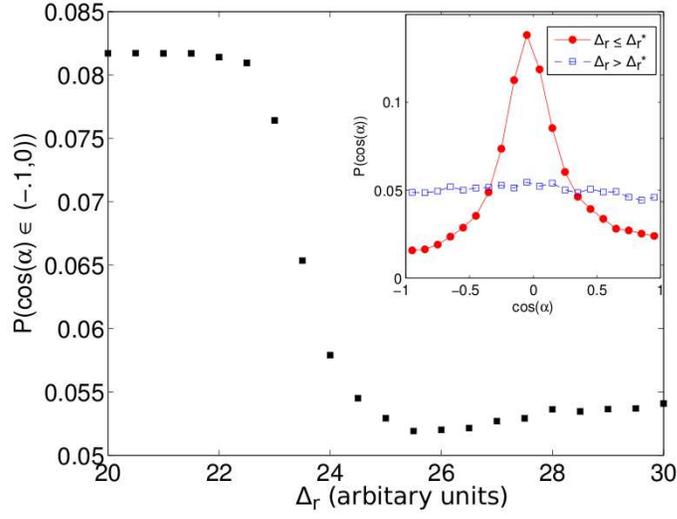}
\caption{(Color Online) The peak value of the $\alpha$ cosines, $P[\cos(\alpha)]$ as a function of radial displacement $\Delta_r$.  Smaller peak values indicate more uniform distributions.  Note the sudden decrease to uniformity as the radial displacement increases.  We take the value after the sudden decrease ($\Delta_r^*=24.5$) as our conservative upper bound on particle contact distance.  Inset: The distribution of $P[\cos(\alpha)]$ for touching $\Delta_r\leq 24.5$ [circles] and non-touching $\Delta_r>24.5$ [squares] particles. }
\label{palpha}
\end{center}
\end{figure}

In order to recover a network linking proximate particles, we need to define which particles interact as neighbors.  However, it is not possible to detect with certainty whether particle pairs are touching - even a nm scale gap between particles would suffice to prevent force transmission between particles.  Therefore we investigate, as a dynamic, statistical  metric of neighbor interactions, how particles move with respect to nearby neighbors.  Recent work \cite{Ellenbroek2006, Katgert2009, Slotterback2009} has demonstrated that under small forcing,  particles in a jammed system predominantly roll or slide past neighbors.  The distribution in the angle, $\alpha$, between the relative displacement vector of a pair of particles and the vector connecting their particle centers is peaked for tangential displacements, indicative of sliding or rolling behavior. For an ensemble of pairs of particles that do not touch and move independently the distribution is uniform, independent of  $\cos(\alpha)$ (for dynamics in three dimensions, for 2D motion it would be independent of $\alpha$).  Thus, by varying the threshold of radial displacements between particles and observing the distribution of the $\alpha$ cosines, $P[\cos(\alpha)]$, we can determine an upper bound for particle contact.  Figure \ref{palpha} plots the peak value of the distribution of $\cos(\alpha)$ vs. radial displacements.  We note that there is a steep decline in the peak value of the distribution (and hence an increasing flatness in the distribution) for increasing radial displacement.  By selecting the displacement immediately preceding the steep decline, we conservatively select a maximum threshold for particle contact.  All particle pairs closer than the specified threshold then make up a proximity network.  

\section{Network Analysis}  
\begin{figure}[!h]
\begin{center}
\includegraphics*[height=.55\textwidth,angle=0,clip]{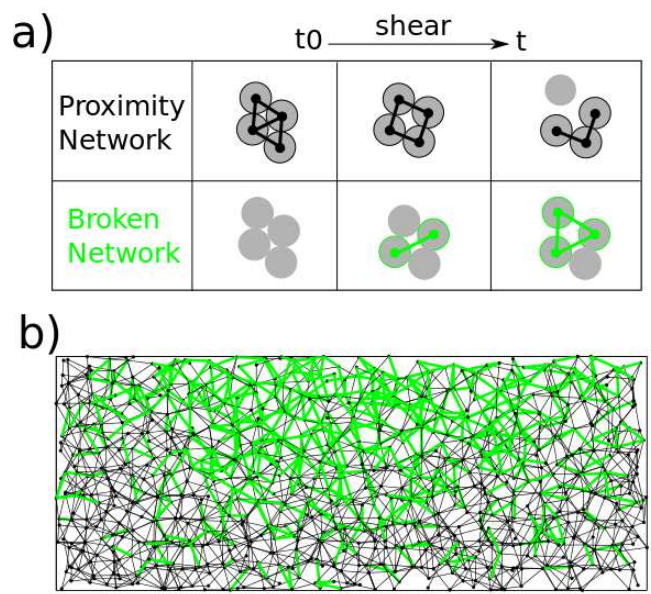}
\caption{(Color Online) (a) A schematic of the definition and growth of the broken-links network.  b) Example cross section of the broken [green/light gray] and persistent [black] networks, plotted in real space.}
\label{brokedef}
\end{center}
\end{figure}

In order to capture the dynamics of the granular flow, we study the time-evolving broken-links network.  A link is considered `broken' relative to some reference frame if the particles it linked in the reference time frame move away from each other some time later (i.e. no longer meet the criterion for assigning contacts).  Thus, the time evolving broken-links network is the collection of all such links (and the nodes/particles they are attached to).   A schematic of this operation is displayed in Figure \ref{brokedef}a.   To lower the occurrence of false rearrangement events,  in order for a link to break, it must be broken in the current frame as well as the subsequent frame. We do not allow links to reform once they have been broken.  

We choose to focus on the broken network because we are primarily interested in the propagation of fracture events.  However, in other applications, it may be instructive to consider the complimentary network of persistent links, defined as the set particles that are linked in the reference frame and remain in contact for all subsequent frames.   Figure \ref{brokedef}b illustrates the complementarity between the broken network and the persistent network.

In addition to considering the experimentally determined broken network, we implement a model of link breakages where the probability of a link breaking is proportional to the average rate of longitudinal strain (elongation) in the direction of the edge. Using the  actual experimentally measured positions of the particles, we begin by choosing a reference time and consider the contact network present at that time.  To consider the effect of increasing shear, we take the experimental particle positions after the shear has been applied and for each edge in the reference network, find $\vec{r'}$ the position of the midpoint between the particle centers.  Thus, $\vec{r'}$ serves as an approximation for the location of the edge in the shear cell, even if the particles have moved away from each other.  

\begin{figure}[!h]
\begin{center}
\includegraphics*[height=0.85\textwidth,angle=0,clip]{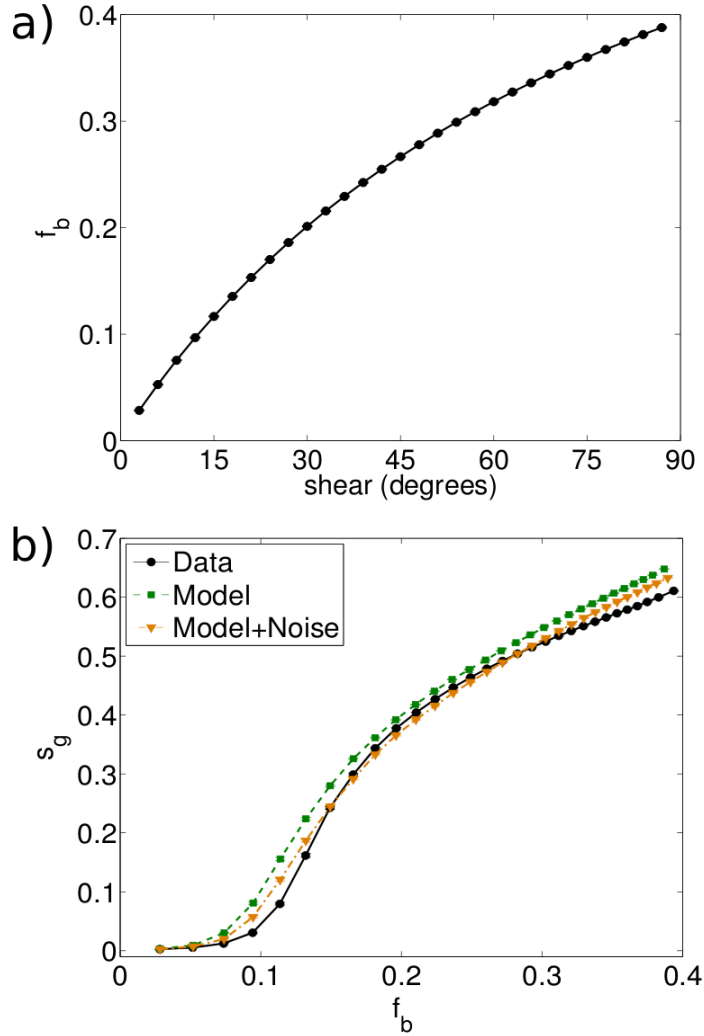}
\caption{(Color Online) (a) The fraction of edges broken as a function of applied shear for the experimental data, averaged over all reference networks. (b) The size of the giant component $s_g$ as a function of the fraction of broken edges for the experimental data [circles], longitudinal strain model [squares], and the longitudinal strain model  with noise ($\eta=0.82$) [triangles].  Errorbars represent the standard error and are smaller than the markers.}
\label{giant}
\end{center}
\end{figure}

Since the dominant average velocity is in the angular direction and uniform as a function of angle, we can approximate $\vec{v}=(v_r,v_\theta,v_z)=(0,r \omega(r,z),0)$, and calculate both the strain rate tensor $\bf{D}$, and the rate of longitudinal strain, 

\begin{align}
\dot{\epsilon} & =\hat{e} \cdot {\bf D} \cdot \hat{e} \\
&= (\hat{\theta} \cdot \hat{e}) \left[ r\frac{\partial \omega}{\partial r}(\hat{r} \cdot \hat{e}) + r\frac{\partial \omega}{\partial z} (\hat{z} \cdot \hat{e}) \right]_{\vec{r'}}
\end{align}
where $\hat{r}$ is taken relative to the center of the disk and $\hat{e}$ is the unit vector from one particle center to the other.  For each amount of shear considered, we define the probability of breaking a link in the contact network, $p$, to be
\begin{equation}
p \propto \Theta(\dot{\epsilon})
\label{meq}
\end{equation}
 where $\Theta$ is the Heaviside step function.  Thus, we only assign probabilities of breaking to edges that have positive longitudinal strain (particles moving apart).  Linear interpolation is used to evaluate the velocity gradients.  The probability of breaking is normalized such that the expected value of the fraction of edges broken, $f_b$, after each amount of shear is applied is equal to the experimentally observed value, as seen in Figure \ref{giant}a.  We also implement the model described above with noise.  For a given amount of shear, a number $M_b$ of the reference edges must be broken in order to match the experimentally observed fraction broken, $f_b$.  We introduce a noise parameter $0\le \eta \le 1$ such that $M_b \eta$, of these edges are broken according to Eq \ref{meq}, and the remaining $(1-\eta)M_b$ edges are broken randomly with uniform probability.  

A particularly useful metric to capture the growth of the broken-links network is the size of its giant component.   The giant component is the largest collection of connected nodes and edges present in the broken network at any given time.   In network theory, it is the
order parameter that signifies a phase transition in the behavior of the system \cite{Newman1999,Callaway2001}.  The size of the giant component, $s_g$, is the fraction of nodes from the reference graph that are in the giant component.  Thus, as shear is applied, one would expect the giant component of the broken network to grow in time. 

The optimal value of $\eta$ is found by maximizing the agreement between the model and the experimental data.  A subset of the reference networks is used to calculate the giant component size as a function of shear, and $\eta$ is tuned to minimize the mean squared difference between the calculated and experimental curves.

We consider one set of constant shear  with a total of 240 frames, corresponding to 717 degrees of rotation of the bottom plate.  The first 210 frames are each taken as a reference frame and define 210 proximity networks.  We allow particle tracks to be considered valid if the largest number of consecutive untracked frames for that particle is less than or equal to one.  This condition is necessary since the experimental data contain some gaps, i.e. particles are not detected in some frames.  For each network, the growth of the broken-links network and the giant component is studied over the subsequent 87 degrees, and we only consider particles whose tracks are valid for those 87 degrees.

We average over all  210 reference networks.  Figure \ref{giant} plots the result of this analysis, with the size (number of nodes) of the giant component (relative to the reference network) as a function of the fraction of edges broken, $f_b$.   It is clear that the fraction of edges broken, $f_b$, plays the role of a tuning parameter, in the same way that an occupation probability is used to tune the behavior of a percolating system.   The data suggest a continuous phase transition in the size of the giant component. One can map the fraction of edges broken to strain, via Figure \ref{giant}a. Thus, one can map the value of $f_b$ at which the giant component forms to a characteristic strain scale corresponding to the onset of global fracture, $s_{char} \sim 15$ degrees.

\begin{figure}[!h]
\begin{center}
\includegraphics*[height=.85\textwidth,angle=0,clip]{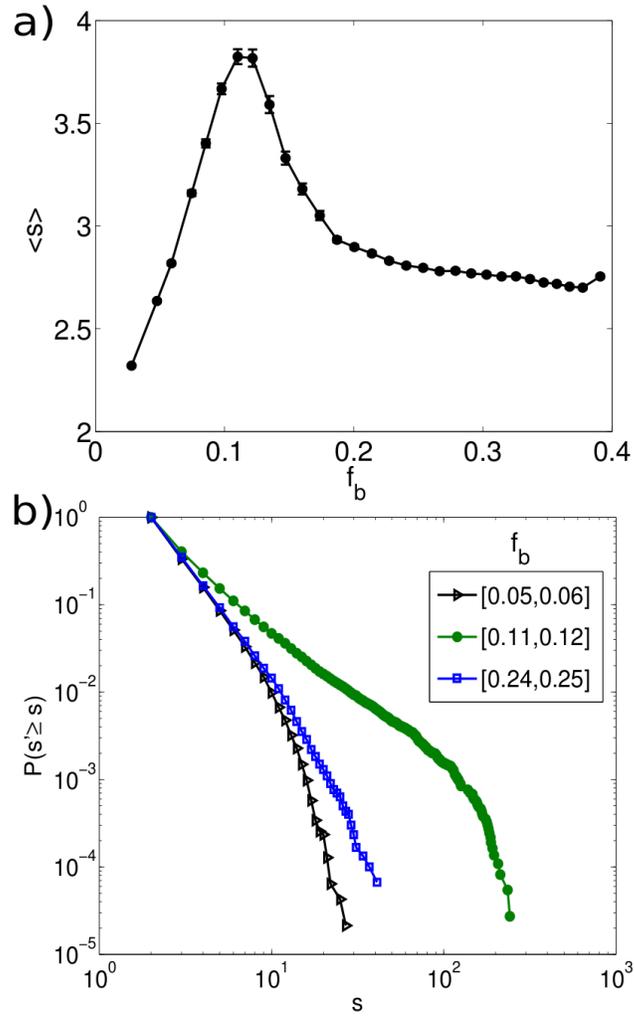}
\caption{(Color Online) (a) The average non-giant cluster size as a function of the fraction of broken edges $f_b$ for the experimental data.  (b) The cumulative size distribution for the non-giant components taken at three different ranges of $f_b$: $0.05 \leq f_b \leq 0.06$ [lower triangles], $0.11 \leq f_b \leq 0.12$ [upper circles], $0.24 \leq f_b \leq 0.25$ [middle squares].  }
\label{mean_s_cumul}
\end{center}
\end{figure}

Continuing the analogy with a percolation phase transition, we can look for a measure associated with a correlation length.  To this end, we plot the average non-giant cluster size as a function of $f_b$ in Figure \ref{mean_s_cumul}a.  Note that mean cluster size peaks approximately in the region in which the transition appears to occur in Figure 
\ref{giant}b.  We can further investigate this transition by looking at the cumulative distribution of cluster sizes at $f_b$ values near this peak.  The cumulative size distribution $P(s' \geq s)$ is the probability that a cluster has a size $s'$ greater than or equal to $s$.  Figure \ref{mean_s_cumul}b plots the cumulative size distribution of clusters for three different ranges in $f_b$: a $f_b$ range well before the transition, a range close to the transition (corresponding to values near the peak of $\langle s \rangle$), and a range well after the transition.  We immediately note that the distribution of cluster sizes for a range near the transition is much heavier tailed than both distributions before or after the transition, as one would expect for a percolation like transition.   

\begin{figure}[!t]
\begin{center}
\includegraphics*[height=.45\textwidth,angle=0,clip]{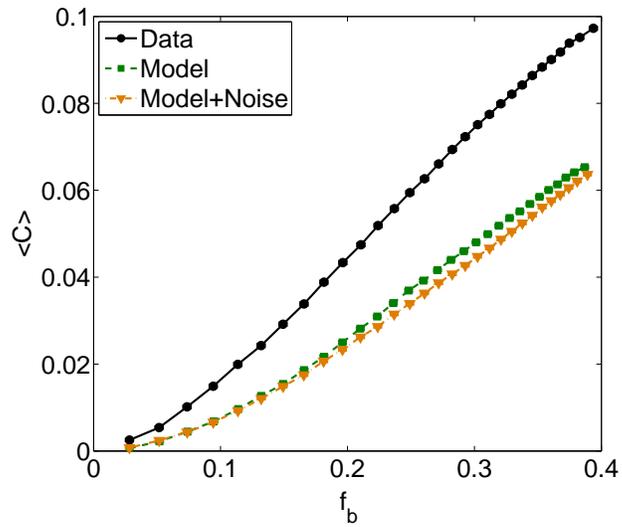}
\caption{(Color Online) The clustering coefficient $C$, as a function of the fraction of edges broken, averaged over all reference networks.  Bins are selected such that each bin contains the same number of data points.   Error bars represent the standard error and are smaller than the markers.  The experimental data [upper circles] are more clustered than both the model [squares] and the model with noise [triangles].}
\label{cluster}
\end{center}
\end{figure}

Additionally, we can investigate the structure of the broken-links network as it grows by computing the network's clustering coefficient, which reflects the extent to which neighbors of a node are themselves neighbors. The clustering coefficient for a node $j$ is defined as 
\begin{equation*}
c_j=\frac{2T(j)}{(k_j)(k_j-1)}
\end{equation*}
 where $k_j$ is the degree of node $j$, the number of neighbors, and the number of triangles $T(j)$ is the number of pairs of neighbors of $j$ that are also linked to each other in the broken link network.  The clustering coefficient measures the fraction of possible triangles centered at node $j$ that are observed.  The network clustering coefficient is simply the average over all nodes of the nodes' clustering coefficients $C=\langle c_j \rangle$ \cite{Saramaki2007}.  Thus, a larger value of $C$ for the broken-links network means that particle contacts tend to break more between nodes whose contacts have broken previously.  Figure \ref{cluster} plots the average clustering coefficient as a function of the fraction of edges broken averaged over all runs.  We note that both the model and the model with noise are less clustered than the data.  Thus, while it appears that the model and model with noise appear to qualitatively reproduce the transition of the giant component, they fail to fully capture the structure of the data.

\section{Discussion} Shear zones and reproducible flow fields are a key feature of granular flows. In order to capture dynamics between particle and bulk scales, we introduced a new, complex networks approach to studying the dynamics of granular flow: the time-evolving broken-links network.  A transition reminiscent of a phase transition in percolation is observed: a giant component of this network develops as shear is applied to this system.  The appearance of this giant component reflects the onset of global fracture in the sense that the breaks in the original proximity network start to become globally connected.  Additionally, we note  a peak in the mean component size in proximity to the observed transition, as well as a heavy tailed distribution in the component size.   This heavy-tailed length scale distribution is reminiscent of the one observed in string-like dynamics \cite{VollmayerLee2006}.  The network approach used here is ideal for studying granular phenomenon that exhibit a diverging length scale of this type.    Our work shows that even in the case of a diverging length scale, it is possible to recover a characteristic strain scale corresponding to the onset of global fracture. 
We hypothesize that this strain scale can be linked to physical features of the system such as the loss of reversibility (this is the focus of ongoing work).   Comparison to a model of link breakages, in which breakage events depend on the average longitudinal strain, highlights the collective nature of dense granular shear flows: the experimental data's broken-links network exhibits a larger clustering coefficient, indicating a more localized fracture than the model.  This suggests that in a granular shear flow, breakage of a link between particles increases the likelihood of breakage of other nearby links.  This further implies that the rearrangement events are cooperative.  Indeed  there is evidence for collective dynamics in dense granular flows, for example collective roll-like dynamics seen in chute flows \cite{Oleh2006} and string like rearrangements in excited dense granular materials \cite{Berardi2010}.  The network analysis provides a characteristic strain scale at which the network topology changes specifically without the need to focus on a particular lengthscale for collective dynamics.  In conclusion, we believe that a dynamic networks approach offers a novel method to quantify granular flows at intermediate scales.

This work was funded by NSF-DMR 0907146.  MH was supported by the Department of Defense (DoD) through the National Defense Science \& Engineering Graduate Fellowship (NDSEG) Program.


%

\end{document}